# Why our human-sized world behaves classically, not quantum-mechanically
## A popular non-technical exposition of a new idea

### C. L. Herzenberg

**Abstract**
This article provides a popular, largely non-technical explanation of how large objects can behave classically while smaller objects behave quantum mechanically, based on the effect of the presence of cosmic expansion velocities in extended objects. This article is intended to provide a more accessible presentation of concepts introduced in earlier papers that address this long-standing enigma in physics.

**Key words**: quantum-classical transition, cosmic expansion, Hubble expansion, uncertainty principle, quantum behavior, classical behavior, Hubble velocity

*What is this about (and why does it matter)?*

Ordinary objects that we deal with in everyday life have specific, definite locations and motions that can be described with considerable precision. We can identify where these 'macroscopic' objects are located and how they are moving quite accurately and predict their motion and future locations in detail, using classical mechanics. This is not the case for extremely small objects, like atoms and elementary particles. These very small objects seem not to be characterized by such precise locations and motions, and instead their locations and motions need to be described by probabilities. The theory that provides this description is called quantum mechanics. Quantum mechanics gives us excellent information about some aspects of the microworld, but it is very different from classical mechanics.[1]

The fact that the everyday reality of our human-sized world is governed by classical mechanics while the microworld of atoms and elementary particles appears instead to be governed by quantum mechanics has been something of a mystery for many years, and may be considered a foundational problem in quantum physics. We understand when we are supposed to use one theory and when we are supposed to use the other, but why there are two different theories of mechanics and why they separate where they do, is not nearly so clear.

Why should there be these two different descriptions, with the objects that we deal with in everyday life being very well described by classical mechanics; while, on the other hand, quantum mechanics is highly successful in describing very small objects, such as atoms and elementary particles? It would seem reasonable to expect that more fundamental laws of physics would be independent of the size of the physical objects that they describe.

It is generally thought that quantum mechanics is the more fundamental theory, and that it must approach classical mechanics in the limit, just as wave optics includes the geometrical optics of light as a limiting case. So why does the world appear classical to us, in spite of its supposed underlying quantum nature? If quantum mechanics were assumed to be universally valid, it would seem that locality would have to be derived.[2,3] It might also be expected that there would be intermediate conditions under which classical physics and quantum physics would overlap or substantially agree, and that quantum mechanics would merge into classical mechanics for objects of intermediate sizes.[2,3]

Many efforts have been made to understand in detail how quantum mechanical descriptions relate to classical descriptions of the behavior of matter, and under what circumstances an object subject to quantum mechanical laws could behave classically. Some of these research efforts have been quite successful, but none so far has produced a really simple and straightforward explanation of the quantum-classical transition.

A remarkable feature of the quantum-classical transition is that it is so robust, - it does not seem to matter what kinds of quantum pieces you put together or how you assemble them, as long as the assembly is big enough, you get this classical behavior. Why exactly is it that when you put lots of small systems together, somehow, as a whole, they behave classically? This also is a fundamental question that we have really not understood very well in the past.[4]

How the state of quantum systems can become effectively classical is of significance for the foundations of quantum physics, and also potentially of importance for many recent applications of practical interest.

So the question that we want to explore is, why can large objects' location and motion be described with precision, whereas the location and motion of small objects cannot be? What makes large objects 'sharp' and small objects 'fuzzy'? And what is it that establishes the threshold between 'large' and 'small', that is, between objects that behave classically and those that behave quantum mechanically?

*Let's visualize what's going on (first!):*

Physical objects seem to have the characteristics of both particles and waves. Classical mechanics describes the particle properties of objects, while quantum mechanics describes the wave properties of objects. (It is not that particles such as electrons are waves, but that the laws of motion in the microworld are wave-like in character.) Wave properties seem to predominate for small objects.

To be more specific, quantum mechanics describes objects in terms of probability waves. Strangely enough, quantum mechanics would describe an object that is completely at rest as a wave extending uniformly throughout space; this leads to a probability distribution that is the same everywhere and is constant throughout all space. Accordingly, an object at rest, behaving quantum mechanically, would be equally likely to be located anywhere.

And in quantum mechanics, an object moving at constant velocity would be described by a wave of uniform amplitude, and would also be equally likely to be located anywhere. Thus, in the simplest quantum mechanical description, an unconstrained object would be equally likely to be found anywhere at all in the universe. But human-sized objects are not found anywhere throughout space with equal probability, they are localized at particular positions. It is clear that quantum mechanical descriptions such as these clash with our experience and do not properly describe the objects of our familiar everyday world. But, why not? Why don't macroscopic objects seem to behave quantum mechanically?

Since ordinary macroscopic objects are localized instead of spread out, it would seem that classical reality could emerge from the substrate of quantum physics if quantum systems were forced into states described by localized wave packets.[5] So, how can objects be localized within a quantum mechanical world? It is possible to localize objects in quantum mechanics, but to do so necessarily causes changes in their motion. In quantum mechanics, any object that is localized would necessarily also be in motion along every direction associated with its localization. In particular, an initially localized object without further constraints would be represented in quantum mechanics by a localized wave packet with waves in motion spreading outward away from its position. A free object in quantum mechanics cannot be localized without also exhibiting this type of spreading or expanding motion away from its position of localization.

If such an expanding motion were already always intrinsically present, could that account for the localization of large objects?

Could we somehow make provision for an expansion of all macroscopic objects enough to permit their localization? Perhaps what is needed is an expansion of space within the volume of every large object?

But, - we know that space is expanding, our whole universe is expanding. Such an expansion might be what saves our classical world from quantum behavior![2,3]

Can it be that the expansion of the universe makes our world of classical physics possible?

*To find out if this could really work, let's continue with an easy approach using the uncertainty principle:*

A very important feature of quantum physics that seems to be truly distinct from classical physics is that in quantum physics there is an intrinsic uncertainty about what is going on. In classical physics, if you are uncertain about something, it's just because there is information that you don't know, or don't know precisely enough.

One approach to exploring the transition between quantum mechanics and classical mechanics would seem to be to examine phenomena that might increase or decrease uncertainties in the location and motion of physical objects. Phenomena that might produce or affect uncertainties in the positions or motions of objects might therefore be considered as potentially having a role in creating the quantum-classical transition.

Everything we deal with directly in our lives is macroscopic and behaves classically; we never get to experience quantum behavior directly, simply because the objects that exhibit it are too small. As a result, most of us think and visualize in terms of the classical behavior of objects, because our life experiences have been with large objects. Accordingly, we try to develop an understanding of quantum behavior from a classical (macroscopic) point of view. Let's now adopt this approach.

There is an easy way of analyzing the quantum behavior of objects, at least in an approximate way, from a human-sized macroscopic point of view. That is by using the Heisenberg uncertainty principle. The Heisenberg uncertainty principle tells you about the uncertainties that must always be present due to wave behavior, without requiring a detailed mathematical analysis of the waves.

What are these uncertainties? We usually think of uncertainties in measurements as due to crudeness or errors, or more generally to lack of knowledge on our part. However, in quantum mechanics there are intrinsic uncertainties that are not associated with the crudeness of our instruments or our sloppiness or our ignorance. These intrinsic uncertainties are present due to the wave nature of matter. (Think about trying to modify a wave pattern. For example, if you have a wave packet and squash it to make it smaller and more localized and more precisely positioned, then the wavelengths will be squeezed smaller and so the motion and momentum associated with the wave packet will change.)

The leading idea of Heisenberg's uncertainty principle is that there will always be some uncertainties, no matter how carefully you control or make measurements on a quantum mechanical object.

What Heisenberg's uncertainty principle tells you about are the relationships between different uncertainties. These are the uncertainties associated with particular characteristics of an object, like its position or its in momentum. (The object's momentum is the product of its mass and its velocity.) Specifically, the uncertainty principle tells you that the product of the uncertainty in an object's position and the uncertainty in the object's momentum cannot be reduced below a particular limit. More specifically, Heisenberg's uncertainty principle states that the uncertainty in the position of an object, times its uncertainty in momentum, must always be larger than or equal to a small fixed universal quantity that is a constant of nature. Heisenberg found that the product of the uncertainties in momentum and position must be greater than a lower limit of $h/4\pi$. (Here, the quantity h is Planck's constant, a very small natural constant characterizing quantum behavior.)[6]

So, Heisenberg's uncertainty principle tells you that you can control an object and make measurements on it as carefully and precisely as you possibly can, but the best that you can possibly do is to have an object whose simultaneous uncertainties in position and in momentum (or velocity) have a product equal to a limiting minimum value.

For clarity, let's suppose for now that we are going to control an object and perform measurements on its motion with the highest precision possible, which is the best precision that Heisenberg's uncertainty principle will allow. Then the product of the uncertainty in momentum times the uncertainty in position will be equal to the smallest possible value, which is h/4π.

Under these circumstances, if the uncertainty in momentum (or velocity) is made smaller, then the uncertainty in position must necessarily become larger; while if the uncertainty in momentum or velocity associated with the object is allowed to be larger, then its uncertainty in position can become smaller. To attain improved precision in the position of an object, there has to be a corresponding decrease in precision in the velocity of the object. Under these circumstances, the uncertainty in an object's position could be reduced in the presence of an increased uncertainty in its velocity.

So, under these circumstances, if the object is permitted to have a sufficiently large uncertainty in momentum (or velocity), then its uncertainty in position could be reduced to as small a value as desired. How small an uncertainty in position would be required so that a quantum object would appear to be behaving like a classical object? To prevent a quantum object from being so spread out that it does not behave like a macroscopic object, we could specify that it should have an uncertainty in position comparable to or small compared to its own size. (Then this object would have only a little bit of quantum fuzziness, not be spread out all over the map.)

Reducing the uncertainty in position to a value small compared to the object's size could take place if the associated uncertainty in velocity were permitted to be adequately large.

But let's return to the question of how there can be uncertainties, and how we can visualize them. How can there be uncertainty in location of an object? In a classical picture, if the object were hopping around with casual disregard for the classical laws of physics, that might perhaps give a visual idea. What about uncertainty in momentum or velocity? Erratic changes in velocity with time would seem to suffice. But since we regard both space and time as aspects of space-time, it would seem that there could be uncertainty in the velocity associated not only with changes in the velocity as a function of time but also with changes in the velocity as a function of space. If the velocity exhibits changes with time or space not accounted for by the customary forces that we consider applied to an object, it would appear that such changes might introduce uncertainties that could affect quantum behavior.

Let's think about what would be implied by such spatial changes in velocity within an extended object, and their effect on the associated uncertainty in position.

If such spatial differences in velocity within an object were to affect quantum behavior so as to cause it to become classical, it would seem that there would have to be an intrinsic velocity spread within objects that would be larger for large objects than it is in small objects, in order for large objects to be able to behave classically. How can there be a spread in velocity that is bigger for large objects and smaller for small objects?

Suppose that there were already an intrinsic spread of velocities within any extended object. Well, there is. At least, there may be, if Hubble's law governing the velocity of expansion in an expanding universe is fully applicable at all distances. Could this play a role in providing an intrinsic uncertainty in velocity for such an object?

So cosmology could be giving us a clue. We live in an expanding universe, in which every point in space is moving away from every other. According to Hubble's law of expansion, the velocity of recession between any two positions in the universe is proportional to the distance between them. The expansion of the universe on cosmological scales has been very well established for many years, and provides the basis for much of our present understanding of cosmology.[7]

Thus, all physical objects in our expanding universe are immersed in an expanding spatial environment. The expansion of the universe is thought to be a property of space itself, so it seems reasonable that the universe may be expanding at all scales of distances, even at very small separation distances. And we know that for cosmological expansion, the recession velocity between positions increases linearly with the distance between the positions. The recession velocities between locations at very small distances from each other would be extraordinarily small, but because the Planck constant that is present in Heisenberg's uncertainty principle is also an extraordinarily small quantity, we need to consider the fact that even very small velocities may be large enough to have a significant effect on positional uncertainties. The spread of Hubble expansion velocities within extended objects might thus affect the uncertainties in the position of all objects.

A feature of the Hubble expansion is that the magnitude of the Hubble expansion (recession) velocity increases with distance. Thus, the larger the object, the greater the spread of Hubble velocities will be within it. Larger objects would have a bigger spread of recessional velocities within their spatial extent, - and this could lead to smaller positional uncertainties for these large objects.

Calculations do appear to indicate that the distinction between quantum and classical behavior may be based in part on the presence of cosmic expansion velocities within the spatial regions occupied by extended objects. The range of magnitudes and directions of the extraordinarily small cosmic recessional velocities within an extended object taken together with the Heisenberg uncertainty relation appears to require an uncertainty in spatial position dependent upon the size of the object. Specifying that such an uncertainty in position be smaller than the size of the object defines a critical size that may provide a fundamental limit distinguishing the realm of objects governed by classical laws from those governed by quantum mechanics.[2,3]

*Let's put some numbers in and check if this explanation can really work:*

So, let's suppose that we have cosmic expansion velocities present within the space occupied by an object. If these Hubble expansion velocities have an effect on the uncertainty in position of the object, how large or small would the uncertainties in position be? If we use the spread in Hubble expansion velocities present in an extended object to calculate the uncertainty in position of that object using Heisenberg's uncertainty principle, we can find out.

(For those who want to delve deeper, an approximate expression that can be derived for the uncertainty in position is: $\Delta x \approx h/(4\pi m H_o L)$, where $\Delta x$ is the uncertainty in position, m is the mass of the object, L is the object's approximate linear size, and $H_o$ is the Hubble constant, which is the proportionality factor that relates Hubble expansion velocities to separation distances.)[2,3]

The results tell us that very small-sized objects with small masses will have large uncertainties in their positions, and so will behave quantum mechanically; however, large objects with large masses will have small uncertainties in position because of the presence of larger Hubble expansion velocities within their extent, so they will behave classically.

How big are the Hubble expansion velocities? At cosmological distances, Hubble velocities are very large, reaching as the velocity of light, but at ordinary human-sized distances they would be miniscule, unmeasureably small. Within an object that is about a centimeter in size Hubble expansion velocities would be of the order of $10^{-18}$ centimeters per second, undetectably small. If such velocities are present, we would have no direct way of knowing about it, as there would be no detectable motion of the object, which would seem to be completely at rest.

Let's put in the numbers for a piece of ordinary matter that is about a centimeter on each side. The Hubble expansion velocities in this object would be minute, of the order of $10^{-18}$ centimeters per second. These tiny velocities would be unobservable, far too small to detect. And yet, the presence of these extraordinarily small Hubble expansion velocities within the volume occupied by this object would be able to reduce the uncertainty in position of this object to roughly $10^{-9}$ centimeter. This uncertainty in position is even smaller than the size of individual atoms, and is extremely small compared to this object's physical size. Such an object would therefore have a very well-defined position. Since any object larger than this would have to encompass an even larger spread of Hubble expansion velocities, we could expect that all objects with sizes in the range of centimeters and larger would be forced by this effect to behave classically.

On the other hand, if we put in the numbers for smaller samples which have smaller masses, we find out that the Hubble velocity spread within them does not reduce the positional uncertainty nearly as much. When we get down to object sizes of about 0.1 mm (or to masses of about a microgram), the uncertainty in position of these objects turns out to be in a range comparable to the size of the objects themselves. Objects in this intermediate range of matter thus might be able to start to exhibit quantum fuzziness about them, so that their behavior could go either way, depending on other environmental influences.[5]

For objects with masses of the order of a microgram we find that this effect therefore leads to a rough limiting threshold between quantum mechanical and classical behavior. This is not unreasonable as an approximate threshold that would provide an upper limit for quantum behavior, as there are other effects such as decoherence effects that can contribute toward bringing about classical behavior in quantum systems.[5]

For objects smaller than this range, the spread of Hubble expansion velocities would necessarily be even smaller. Thus, the positional uncertainty would be even larger than the size of the object, and that would rule out classical behavior. That is, for objects smaller than about a microgram, the effect on positional uncertainty consequent to cosmic expansion within the object's volume would be inadequate to allow localization of such an object. Accordingly, any such smaller sample of matter would be expected to exhibit quantum mechanical behavior as an entire object, unless its behavior is sufficiently affected by other environmental influences to bring about classical behavior.

We need to bear in mind that these are only rather rough estimates, but they seem to be at least roughly in a range to provide an upper limit on quantum behavior and account for the quantum-classical transition between true quantum behavior and the classical behavior of the macroscopic objects that we deal with in everyday life. And this effect will systematically provide for the classical behavior of all large objects.

Observationally, there is not a sharply-defined dividing line or border between macroscopic objects whose physical behavior can be described by the rules of classical mechanics, and the microscopic and submicroscopic objects that usually require quantum descriptions. Effective classical behavior can extend down to the level of medium-sized molecules. (Most molecules except the very small ones seem to have well-defined spatial structures.) Generally speaking, the separation seems to occur in the vicinity of $10^{-6}$ centimeters. At smaller distances, around $10^{-7}$ centimeters, clear quantum effects like quantum tunneling occur reproducibly. At larger sizes, there are viruses as large as $10^{-5}$ centimeters. So, effectively around $10^{-6}$ centimeters is very roughly where the domain fully governed by quantum mechanics seems in practice to begin.

The effect of Hubble expansion velocities would not by itself bring a quantum-classical transition down quite to these lower levels; rather, it sets a somewhat higher limit above which quantum behavior would be unable to occur. But the presence of Hubble expansion velocities has the powerful effect of setting a threshold above which no quantum behavior would take place for individual objects. Some degree of classical behavior observed at smaller distances may result from decoherence in interactions with the environment.

We have noted that one remarkable feature of the quantum-classical transition is that it is so robust, - it does not seem to matter what kinds of quantum pieces are put together or how they are assembled; as long as the assembly is big enough, it results in classical behavior.[4] Why exactly is it that when you put lots of small systems together, somehow, as a whole, they behave classically? The effects of Hubble expansion could provide the needed explanation. So another important aspect of the contribution of Hubble expansion velocities to the quantum-classical transition is that, since it depends only on size and mass, not on detailed composition or other effects, this effect can explain systematically why the transition from quantum to classical behavior is so robust.

*So, we conclude. . . . .*

Thus, it appears that the presence of Hubble expansion velocities may be introducing uncertainties in the motion of large objects that are sufficiently large that, in accordance with Heisenberg's uncertainty principle, the uncertainties in position become small enough that these objects will appear to behave classically. Hubble expansion velocities present within an extended object seems to have the effect of setting a threshold above which quantum behavior for the entire object will not take place.

So, it seems that the presence of cosmic expansion velocities in matter may be what makes our world of classical physics possible. The reason that we live in a classical world rather than in a quantum world may well be because we live in an expanding universe.